\documentclass[12pt]{article}
\usepackage{float}
\usepackage{graphicx}
\usepackage{epstopdf}
\DeclareGraphicsExtensions{.pdf,.png,.jpg} 
\graphicspath{{figure/}}                   

\usepackage{tabularx}
\usepackage{amssymb}
\usepackage{amsthm,lineno}
\usepackage{multirow}

\usepackage{float}
\usepackage{indentfirst}
\usepackage{cite}
\usepackage{amsmath,amsthm,amssymb}
\usepackage{bm}
\usepackage{bbm}
\usepackage{wrapfig, enumitem}
\usepackage[colorlinks = true,citecolor=blue]{hyperref}
\usepackage{color}
\usepackage[round]{natbib}
\usepackage{algorithm}
\usepackage{algorithmic}
\usepackage{graphicx}
\usepackage{booktabs}
\usepackage{longtable}
\usepackage{hyperref}
\usepackage{natbib}
\usepackage[a4paper,margin=2.5 cm]{geometry}
\usepackage[toc,page]{appendix}
\usepackage{comment}
\usepackage{diagbox}
\usepackage{makecell}
\usepackage{threeparttable}
\usepackage{multirow}
\usepackage{titlesec}
\usepackage{booktabs}
\usepackage{subfig}
\usepackage[toc,page]{appendix}
\titlespacing*{\section}{0pt}{3pt plus 3pt minus 1pt}{3pt plus 3pt minus 1pt}
\titlespacing*{\subsection}{0pt}{3pt plus 3pt  minus 1pt}{3pt plus 3pt minus 1pt}

\title{Chitchat with AI: Understand the supply chain carbon disclosure of companies 
worldwide through Large Language Model}
\author{Haotian Hang$^{1}$, Yueyang Shen$^{2}$, Vicky Zhu$^{3}$, Jose Cruz$^{4}$ , Michelle Li$^{3}$ \\
        \small $^{1}$University of Southern California, Los Angeles, CA, USA \\
        \small $^{2}$University of Michigan, Ann Arbor, MI, USA \\
        \small $^{3}$Babson College, Wellesley, MA. USA\\
        \small $^{4}$University of Connecticut, Connecticut, CT, USA
}
\date{} 
\usepackage{float}
\begin{document}

\maketitle
\begin{abstract}
In the context of global sustainability mandates, corporate carbon disclosure has emerged as a critical mechanism for aligning business strategy with environmental responsibility. The Carbon Disclosure Project (CDP) hosts the world’s largest longitudinal dataset of climate-related survey responses, combining structured indicators with open-ended narratives, but the heterogeneity and free-form nature of these disclosures present significant analytical challenges for benchmarking, compliance monitoring, and investment screening. This paper proposes a novel decision-support framework leveraging large language models (LLMs) to assess corporate climate disclosure quality at scale by developing a master rubric that harmonizes narrative scoring across 11 years of CDP data (2010–2020), enabling cross-sector and cross-country benchmarking. By integrating rubric-guided scoring with percentile-based normalization, the method identifies temporal trends, strategic alignment patterns, and inconsistencies in disclosure across industries and regions. Results reveal that sectors such as Technology and countries like Germany consistently demonstrate higher rubric alignment, while others exhibit volatility or superficial engagement, offering insights that inform key decision-making processes for investors, regulators, and corporate environmental, social, and governance (ESG) strategists. The proposed LLM-based approach transforms unstructured disclosures into quantifiable, interpretable, comparable, and actionable intelligence, advancing the capabilities of AI-enabled decision support systems (DSSs) in the domain of climate governance.
\end{abstract}


\section{Introduction }

In a world shaped by economic prosperity and booming trade, yet overshadowed by accelerating environmental crises and growing public scrutiny, companies now face an unprecedented dual mandate from investors and regulators: generating shareholder value while demonstrating robust environmental accountability. International policies like the Paris Agreement and frameworks like the Task Force on Climate-related Financial Disclosures (TCFD) have raised expectations for transparent climate reporting. These international agreements add complexities to companies to fulfill climate responsibilities while generating financial success to sustain and grow their global portfolio. Companies around the world face challenges to balance competing demands from vested stakeholders while navigating the constraints imposed by sustainability requirements. 

The CDP dataset, which covers responses from thousands of major companies in many different economies from 2010 to 2020, features questions and responses aiming to collect sufficient information on climate-related performance disclosures. As the largest ongoing survey-based platform for company carbon disclosure, the dataset comprises hybrid data formats, including both structured answers and open-ended textual responses that detail emission reduction strategies, supply chain risks, and governance frameworks. The CDP format uniquely suits the textual analysis as it offers a standardized framework combined with qualitative richness~\citep{cao2025carbonchat, armbrust2022deep}.
Although previous work on CDP data shows an increasing trend, indicating that companies are becoming more transparent about their environmental responsibilities~\citep{blanco2021supply}, there is no comprehensive framework to consistently evaluate the effectiveness and progress of these companies' efforts. The textual data in CDP contains rich information, which allows us to study how the companies' responses relate to their economic growth and environmental sustainability. On the other hand, the heterogenetic nature deriving from the free-form response and the explicit time dimension involved also requires extra care to handle and interpret the data.

The global and longitudinal scope of CDP disclosures enables cross-sectional comparisons and tracks sustainability communication trends over time. However, the variability in narrative tone, depth, and language calls for sophisticated NLP techniques, particularly those capable of semantic embedding and strategic alignment assessment. Concurrently, LLMs have emerged as a transformative force capable of interpreting complex, unstructured corporate narratives at scale~\citep{achiam2023gpt,team2023gemini,liu2024deepseek}. These cognitive copilots have compensated for limited human thought throughput and attention span through rapid token generation, seamless thought continuation, streamlined planning, effective summarization, and holistic insight integration.

This convergence of high environmental stakes and AI capability presents a unique opportunity. As firms increasingly adopt voluntary climate disclosure protocols, these firms rely on semi-structured, free-form narratives to characterize their effort and stance in response. These documents contain rich semantic content, but their heterogeneity poses challenges to analysis. Namely, the variation across companies, industries, and time makes them challenging to synthesize and analyze this fragmented information with traditional tools. LLMs offer a scalable approach to extract strategic intent, tone, and quality from climate-related texts. 

To advance the analysis of global carbon disclosure practices, we focus on analyzing companies' business strategies within the CDP survey for their carbon emission reduction. Specifically, we use LLMs to understand their textual responses. The LLM-based approach acts as a structured assessment pipeline, which allows us to design prompts with an explicit rubric and receive a score for any response from a company~\citep{Mizumoto2023, Zhang2024AES}. 

Our manuscript is structured as follows. Section~\ref{sec:literature} provides a literature review in which we critically evaluate existing approaches to analyzing corporate climate disclosures, ranging from traditional statistical models and classical natural language processing (NLP) techniques to recent advancements in LLMs. This discussion highlights both methodological gaps and emerging opportunities in AI-assisted sustainability analytics. Section~\ref{sec:data} describes the CDP dataset, outlining its scope, structure, and temporal coverage (2010–2020), complemented by visualizations that illustrate industry participation trends and geographical variation. Section~\ref{sec:methodology} introduces our methodological contribution, presenting the design of an LLM-based scoring framework. This section details the rubric-guided evaluation strategies, the construction of a master rubric for temporal re-alignment, and the statistical analysis of scores for benchmarking and comparison. Section~\ref{sec:policyinterpretation} turns to business insights and policy interpretation, leveraging the derived scores and percentile rankings to analyze disclosure trends across sectors and countries, while also linking disclosure dynamics to international climate policy milestones such as the Paris Agreement. Here, we further assess cross-sector and cross-country alignment using correlation and time-series evaluation. Finally, Section~\ref{sec:conclusion} concludes the paper by summarizing the core findings and demonstrating how our framework advances DSS functionalities, with implications for corporate managers, regulators, and investors.

\section{Literature review}\label{sec:literature}
The intersection of environmental sustainability, corporate disclosure, and artificial intelligence (AI) is increasingly relevant to the design of modern DSSs. As firms increasingly utilize open-ended narratives in their climate disclosures, traditional statistical tools have limitations in addressing nuance, strategic alignment, and credibility. This literature review summarizes methodological developments in traditional econometrics, classical NLP, and recent progress in LLMs. Specifically, we evaluated how these tools enable scalable, explainable, and semantically detailed assessments of corporate climate strategies, which are beneficial for informed decision-making by investors, regulators, and firms. The study is framed within the DSS research stream, focusing on methods that improve transparency, benchmarking, and the extraction of strategic insights from unstructured disclosure data. The rest of this section reviews approaches to CDP and corporate disclosure, from traditional methods and early NLP methods to recent LLM-based frameworks. We map these methods to decision contexts, compare their performance, and highlight research gaps that motivate improved DSSs. 

\subsection{Traditional approaches to CDP and corporate disclosure}

Early research into corporate climate disclosures predominantly employed quantitative models such as Ordinary Least Squares (OLS), logistic regression, and descriptive statistics. These studies focus on observable indicators, such as disclosure frequency, participation rates, and emissions data. For example, \citet{cohen2023institutional} demonstrated that institutional investor pressure significantly increases the likelihood of CDP participation and is correlated with reduced emissions. Similarly, \citet{DAmico2016} found that governance and firm size influenced the propensity to disclose.

Despite offering valuable insight, these models lack the granularity needed to interpret qualitative narratives. Moreover, they struggle with the heterogeneity and complexity inherent in cross-sectoral and multi-year disclosure datasets. These limitations become pronounced when companies selectively present data or engage in greenwashing, a challenge that has motivated the use of more advanced text analysis techniques~\citep{bingler2024cheap}.

\subsection{Classical NLP and early machine learning approaches}

Before the emergence of deep learning and transformer-based models, natural language analysis of open-ended climate disclosures, such as those found in the CDP dataset, relied heavily on unsupervised topic modeling and shallow supervised classifiers ~\citep{li2010textual}. The most common techniques included Latent Dirichlet Allocation (LDA)~\citep{blei2003latent} to discover dominant topics, Support Vector Machines (SVMs)~\citep{cortes1995support}, and logistic regression to classify and analyze sentiment. These early models provided valuable initial insights; for example, LDA could surface recurrent themes such as “supply chain risk” or “renewable energy investment” from large corpora of text, enabling simple trend analysis across industries or years~\citep{tidy2016role}. Similarly, SVMs were used to predict the likelihood of disclosure or to categorize responses according to predefined ESG labels.

Despite their contributions, these approaches faced major limitations that restricted their utility in modern decision support systems. First, they lacked contextual understanding: both LDA and SVMs treated words as independent features, ignoring sentence structure and narrative tone, which is critical in climate-related disclosures where companies often blend factual content with aspirational language. Second, they relied heavily on manual feature engineering, with preprocessing pipelines requiring cleaning, stop-word removal, stemming, and vectorization, which reduced scalability and increased the risk of bias. Third, they suffered from poor multilingual support, as most traditional models assumed monolingual input and were ineffective for analyzing CDP’s global multilingual data without extensive translation efforts.

As a result, while these methods were useful for basic summarization, they lacked the capacity to provide information relevant to decision-making, such as assessing sincerity, specificity, or alignment with corporate climate strategies. This methodological ceiling ultimately led to the growing adoption of transformer-based LLMs, which better serve the complex interpretive needs of modern decision support system applications.

\subsection{Emergence of LLMs in climate disclosure}

The introduction of transformer-based models, such as BERT ~\citep{devlin2019bert}, RoBERTa ~\citep{liu2019roberta}, and GPT-3 ~\citep{brown2020language},  GPT-4 ~\citep{achiam2023gpt}, has significantly advanced the analysis of complex open-ended survey data by enabling semantic understanding of corporate narratives and supporting both prompt-based scoring and embedding-based clustering. For instance, \citet{gweon2023automated} found that BERT outperformed SVMs in survey response classification as training data increased, while \citet{mellon2024important} demonstrated that GPT models could categorize open political survey responses with near-human accuracy. Similarly, \citet{petukhova2024text} showed that LLM embeddings outperformed traditional clustering approaches in thematic analysis. Collectively, these studies highlight the scalability, cross-lingual robustness, and contextual precision of LLMs, traits that are especially valuable in the context of CDP analysis, where responses vary widely across countries, years, and sectors.

\subsection{LLM-based scoring frameworks}

Delegating humans to evaluate all the open responses is usually very costly (both time and monetary-wise) due to our limited attention span, working memory, thought generation throughput, and context window. The high-throughput of modern computing infrastructure and innovations in algorithmic design — for example, parallelized decoding methods and speculative decoding~\citep{leviathan2023fast} have enabled LLMs to take in millions of tokens in seconds and generate at hundreds \citep{haghighat2025sota} or thousands \citep{qian2024bass} of tokens per second reliably. 
The high throughput and the high success rate from the next token generation have unlocked a wide range of new possibilities. However, realizing these capabilities responsibly requires the introduction of additional guardrails, oversight, and governance mechanisms. One possible instance is the use of domain-specific prompting, which ensures that raw generation is grounded in factual accuracy and task relevance~\citep{wei2022chain,wang2022self, wang2022self2}. 

Using a language model and template evaluation software to guide and automate the evaluation process has been shown to be effective ~\citep{Leng2023}, and strikes a balance between comprehensive human evaluation and scalable automated evaluation ~\citep{gu2024survey}. Common open source frameworks as of writing include: open-ai-eval, RAGAS ~\citep{es2024ragas}, deepeval ~\citep{Niasojasingarayar_2025}. However, they are primarily targeted at open-ended question answer generation and are not tailored towards the unique challenges of interpreting heterogeneous longitudinal tabular data. 

Besides the automation through software, another flexible pathway for automating decision support is the use of rubric-guided LLM scoring. 
Inspired by educational assessment and policy evaluation, this approach leverages large language models to assign scores to narrative responses based on structured rubrics. For example, \citet{Zhang2024AES} showed that GPT-4 surpasses fine-tuned BERT in rubric alignment and score reliability, while \citet{Mizumoto2023} applied GPT-3 to TOEFL essay scoring and demonstrated human-level agreement. Similarly, \citet{Lee2024CoT} reported that chain-of-thought prompting improves accuracy and consistency in scoring tasks by more than 13\%. In the context of environmental disclosures, \citet{bingler2024cheap} employed GPT-based scoring to detect ``cheap talk" and greenwashing, finding a strong alignment with human annotations. Collectively, these scoring approaches advance the objectives of DSSs by enabling explainable and reproducible evaluation frameworks.


\subsection{Mapping methods to decision contexts}
The power of disclosure analytics lies not just in methodological accuracy, but in decision-making applicability. Within the DSS domain, the ultimate objective of deploying LLMs or statistical models is to generate actionable insights across stakeholder roles, from corporate management to regulatory oversight. This section explores how different methodologies contribute to specific decision support use cases in the context of climate-related disclosures, based on CDP data (Table~\ref{tab:DSS}).

We align key decision actors with methods that enable interpretability, benchmarking, and forward planning—three pillars of DSSs. As LLMs gain traction, their ability to summarize intent, assess sincerity, and project future compliance becomes central to supporting climate-resilient strategies.

\begin{table}[h]
\centering
\resizebox{\textwidth}{!}{%
\begin{tabular}{lll}
\hline
\textbf{Stakeholder} & \textbf{Decision Use Case} & \textbf{Methodological Contribution} \\
\hline
Corporate Managers & Strategic ESG planning & LLM scoring, rubric alignment \\
Investors & ESG screening &  Sentiment analysis \\
Regulators & Policy auditing, compliance & Greenwashing detection, scoring audits \\
Researchers & Benchmarking disclosure quality & Clustering, panel analysis, score tracking \\
\hline
\end{tabular}%
}
\caption{Decision contexts enabled by disclosure analytics methods}
\label{tab:DSS}
\end{table}

LLMs, particularly when guided by explicit rubrics, enhance \textit{cross-functional decision-making}. For example, investors benefit from score clustering to detect strategic leaders vs laggards, while regulators can prioritize audits based on high-volume disclosures lacking specificity. The ability to \textit{map intent to action}, not just text to score, defines the methodological frontier for DSSs in sustainability analytics.

\vspace{1em}

\subsection{Comparative evaluation of methods}
 
As AI models proliferate in climate analytics, selecting the appropriate method for a DSS application requires careful evaluation of \textit{interpretability}, \textit{scalability}, and \textit{decision relevance}. In this section, we provide a structured comparison of the major methodological approaches discussed, benchmarking them against criteria most critical for decision support environments. Specifically, we compare six categories of methods commonly applied in climate disclosure analytics, ranging from traditional regression models and topic modeling to advanced LLM-based frameworks. The evaluation is organized around five key dimensions: \textit{interpretability}, which assesses how easily a human decision-maker can understand the model’s output; \textit{scalability}, which considers whether the method can handle thousands of firms across time and geography; \textit{accuracy}, which measures how effectively the method captures the intended signal, such as disclosure sincerity; \textit{decision relevance}, which examines whether the approach generates insights useful for planning, auditing, or investment; and \textit{multilingual support}, which evaluates the ability of the method to process global datasets that include non-English responses. This structured framework provides a comprehensive lens for assessing the suitability of different analytical tools in the context of climate governance and DSS applications.

\begin{table}[h]
\centering
\resizebox{\textwidth}{!}{%
\begin{tabular}{lccccc}
\hline
\textbf{Method} & \textbf{Interpretability} & \textbf{Scalability} & \textbf{Accuracy} & \textbf{DSS Relevance} & \textbf{Multilingual} \\
\hline
OLS/Logistic Regression & $\bigstar\bigstar\bigstar$ & $\bigstar\bigstar\bigstar$ & $\bigstar$ & $\bigstar$ & $\bigstar$ \\
LDA/SVM & $\bigstar\bigstar$ & $\bigstar\bigstar\bigstar$ & $\bigstar\bigstar$ & $\bigstar\bigstar$ & $\bigstar$ \\
Fine-tuned Transformers & $\bigstar\bigstar$ & $\bigstar\bigstar$ & $\bigstar\bigstar\bigstar$ & $\bigstar\bigstar\bigstar$ & $\bigstar\bigstar$ \\
Few/Zero-shot LLMs  & $\bigstar$ & $\bigstar\bigstar\bigstar$ & $\bigstar\bigstar\bigstar$ & $\bigstar\bigstar\bigstar$ & $\bigstar\bigstar\bigstar$ \\
LLM (Embedding) & $\bigstar\bigstar$ & $\bigstar\bigstar\bigstar$ & $\bigstar\bigstar\bigstar$ & $\bigstar\bigstar\bigstar$ & $\bigstar\bigstar\bigstar$ \\
\hline
\end{tabular}%
}
\caption{Comparative evaluation of methods used in CDP disclosure analysis, High: $\bigstar\bigstar\bigstar$, Medium: $\bigstar\bigstar$, Low: $\bigstar$}
\end{table}

The trade-off between \textit{interpretability and scalability} is central to DSS model selection. Traditional models are easy to explain but lack depth, whereas LLMs offer nuanced analysis at scale—though at a cost to transparency. 
Scoring-based methods, namely using LLM as a judge, strike a balance, allowing for quantifiable and interpretable comparison of disclosure narratives while retaining sectoral explainability by inspecting proper LLM reasoning trajectories.
In DSS contexts where \textit{auditing, benchmarking, or public accountability} is paramount, rubric-guided scoring presents a robust path forward, especially when results can be interrogated and linked to financial metrics or sector-level trends.

\subsection{Research gap and relevance to DSSs}

Despite the substantial progress in leveraging AI models—particularly LLMs—for climate disclosure analysis, several critical research gaps remain, especially in relation to their integration into actionable DSSs.

First, while prior work has demonstrated the value of rubric-guided LLM scoring, \textbf{few studies offer a unified framework that connects these semantic insights to quantifiable business metrics or investment outcomes}. Most LLM applications in climate disclosure remain descriptive, lacking mechanisms to translate qualitative narrative scores into indicators that can inform financial planning, supply chain strategy, or ESG risk assessment. This limits their practical value to corporate executives, investors, and policymakers who require grounded, business-relevant analytics.

Second, the temporal inconsistency in climate disclosure formats (e.g., evolving CDP questionnaires) and language variation across companies poses a significant challenge to traditional statistical or rubric-based scoring approaches. \textbf{Few existing works have addressed rubric alignment across time or sectors}, leading to validity concerns when comparing firm strategies longitudinally. Without normalization, rubric-based LLM scoring may unfairly bias scores across different timeframes or regional templates.

Third, although some studies (e.g., \citet{Chuang2025Greenwashing}; \citet{pathak2025rubric}) apply LLMs to detect greenwashing or improve scoring transparency, \textbf{there is a lack of rigorous evaluation protocols to benchmark the consistency, fairness, and reproducibility of LLM-generated scores}. As \citet{shen2023large} pointed out, LLMs can exhibit variance in judgment quality, especially when prompt engineering or dataset partitioning is suboptimal. This lack of reliability limits their use in sensitive regulatory or compliance contexts.

Our study addresses these key limitations through the following innovations:

\begin{itemize}
    \item \textbf{Master Rubric Framework}: We create a time-agnostic scoring rubric that harmonizes semantic criteria across 11 years of CDP data. This rubric ensures that corporate disclosures from different years and sectors are evaluated under a unified lens, enabling robust cross-temporal benchmarking.
    
    
    \item \textbf{Percentile-Based Ranking}: To supplement raw rubric scores, we calculate intra-year percentiles for each company. This normalizes inter-year variation and provides clearer insight into how firms evolve relative to peers rather than in isolation.
    
    \item \textbf{Validation via Rank Correlation}: We employ Kendall’s $\tau$ coefficient to validate consistency between yearly and master rubric scores. This statistical check ensures that rubric unification retains rank order fidelity, enhancing credibility for DSS deployment.
    
    \item \textbf{Alignment with DSS Objectives}: Our framework directly supports decision-making use cases:
    \begin{itemize}
        \item For \textit{investors}, it enables ESG portfolio screening through percentile-based strategic alignment.
        \item For \textit{regulators}, it flags inconsistent or superficial disclosures for potential audit.
        \item For \textit{corporate managers}, it provides a data-driven basis for refining ESG communications and comparing performance against peers.
    \end{itemize}
\end{itemize}

In summary, our approach bridges the methodological innovation of LLM scoring with the applied utility of decision support, offering a scalable and statistically grounded framework for interpreting climate disclosures. This positions LLM analytics not just as tools for summarization or classification but as engines for strategic foresight, regulatory benchmarking, and long-term sustainability planning.

\section{Data}\label{sec:data}

\begin{figure*}[h]
\centering
\includegraphics[scale=1]{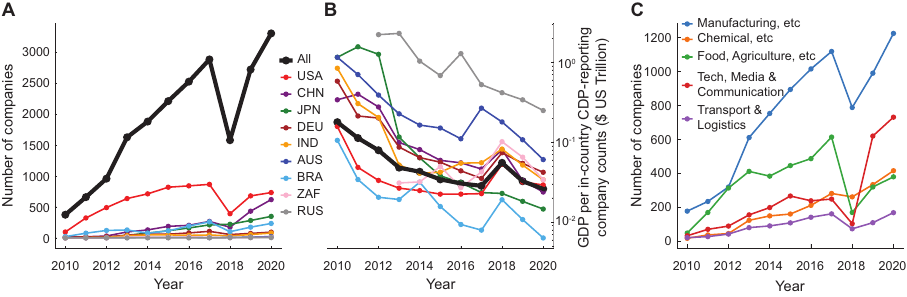}
\caption[n]{\footnotesize \textbf{Summary of supply chain firms participated in Carbon Disclosure Project (CDP)  from year 2010 to year 2020.}  \textbf{A.} Number of firms within each country participated in CDP \textbf{B.} Gross Domestic Product (GDP) divided by the number of companies within each country. \textbf{C.} Number of companies in different industry sectors. \footnote{Different sectors are not mutually exclusive.}
}
\label{fig:company_num}
\end{figure*}

\begin{figure*}[h]
\centering
\includegraphics[scale=1]{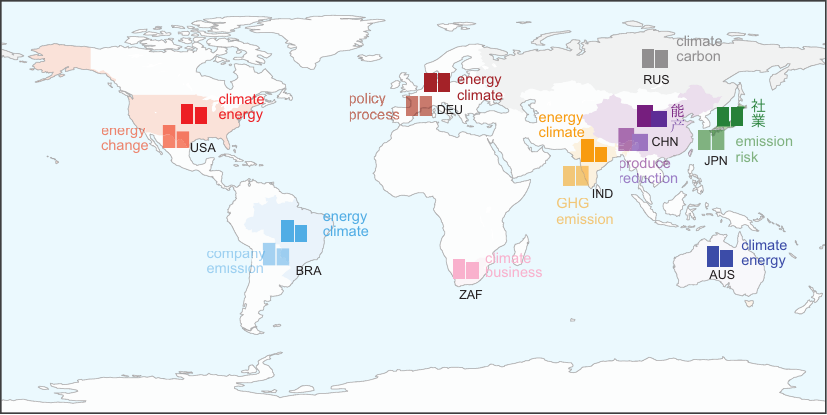}
\caption[n]{\footnotesize \textbf{Word frequency of table changes over time.} The two most common words/characters in the business strategy sector of the CDP survey of companies within certain countries in 2010 (left) and 2020 (right). The height of bar shows the relative counts between the first and second most common word. Note that Russia, South Africa, and Australia do not have data from 2010 and only 2020 data is shown. 
}
\label{fig:word_frequency}
\end{figure*}

CDP is the largest and most comprehensive data source that records companies’ efforts on carbon disclosures at the global level \citep{blanco2021supply, cohen2023institutional}. It captures companies' responses in different industries and regions in multiple-choice, restricted-response, and open-ended forms. Fig.~\ref{fig:company_num}A,B gives a summary of the participation of supply chain companies in each country based on their carbon disclosures and the gross domestic product (GDP) between 2010 and 2020. We see a general trend of increasing participation in CDP within each country as well as in different sectors. Moreover, GDP per reporting company counts decreases, showing that an increase in CDP reporting companies is not merely an effect of the growth of the world economy. Another interesting finding is that there is a sharp decline in CDP participation in 2018, led by the USA, China, and Brazil. Correspondingly, except for the chemical sector, the same decline also appears among other sectors such as manufacturing, food, etc. (Fig.~\ref{fig:company_num}C). To understand these anomalies in carbon disclosure and gain more business insight, we focused our study on the CDP business strategy-related questionnaires sector from 2010 to 2020 for five key reasons.

First, we are interested in analyzing the contexts related to the business strategy, as it is meaningful to see how environmental responsibilities guide companies' business decisions and impact market value and vice versa \citep{matsumura2014firm, Matisoff2013, Matisoff2023}. Second, sufficient consistency is present in the questionnaire structure and responses received in the business strategy area from participating companies in the period of 2010–2020. 
Third, disclosure data during this period provides a decade-long basis for evaluating whether companies exhibit stable reporting patterns and detecting meaningful changes over time, which offers recency and sufficient data quality to support reliable longitudinal analysis. Fourth, we believe that the year 2015 to 2020 is an important window, because they cover many significant events and public policies, such as the Paris agreement in 2015, the US presidential election in 2016, the IPCC special report on global warming in 2018, and the COVID pandemic from 2019. We want to see how the companies are affected and navigate their business strategy while maintaining their global sustainable image under these extreme events. Therefore, most of the analysis we are interested in should be reflected in this time. Fifth, the CDP data structure has changed over the years, especially after 2012. There is a major expansion of the climate scope, which includes wider environmental themes. After 2018, CDP introduced the concept of carbon price and a new reporting framework (we also note that this change may be attributed to the decrease in the number of reporting companies, as shown in Fig.~\ref{fig:company_num}), which yields a more context-rich questionnaire and provides room for more narrative responses. This makes the survey content rich, but also introduces some challenges in analysis. For example, the responses may be disclosed in different languages and in different lengths, etc. In Fig.~\ref{fig:word_frequency}, we show a global comparison of a bar graph representing the change in frequency of the two most common words in the CDP business strategy sector from 2010 (left) to 2020 (right). With a broader representation of companies, many responses are furnished in various foreign languages. The traditional statistical approach in text mining focuses on sentiment analysis through the word count feature engineering before applying a machine learning algorithm \citep{tidy2016role, blanco2021supply}. 
These traditional NLP methods, such as  N-grams~\citep{shannon1948mathematical}, inverse-document frequency (tf-idf)~\citep{sparck1972statistical}, and SVMs ~\citep{cortes1995support}, struggle to align such vocabulary into a unified English space. To supplement the traditional way of analyzing text through word count and a sequential set of processing steps, we propose a modern heuristic that involves LLMs, which also adds novelty to this dataset case study while aligning with the best practices in the NLP space.

\section{Proposed rubric-based LLM grading framework}\label{sec:methodology}
In this section, we introduce a heuristic that quantitatively analyzes the open-ended survey data on the business strategy section within the CDP questionnaire. To begin with, we briefly review the background for LLMs, which sets the stage for the LLM-based evaluation on the text input (Fig.~\ref{fig:scoring_naive}). Capitalizing on these foundations, we introduce a scheme that uses LLMs to create an aggregated rubric to explain the scoring mechanism (Fig.~\ref{fig:our-workflow}).

\subsection{LLM background and tabular data} \label{sec:llm}
Large-scale pretraining has endowed LLMs with broad, generalistic capabilities~\citep{achiam2023gpt,touvron2023llama}, enabling them to follow instructions across diverse domains—including tasks involving more structured formats such as tabular data~\citep{fang2024large}. Since CDP disclosure data is inherently collected from a tabular form and contains various open-ended texts scattered across the table entries, LLMs are suitable to ingest the information in this context. The robust instruction-following and generalization abilities in LLM facilitate effective out-of-the-box usage without requiring task-specific fine-tuning~\citep{wen2024supervised,ouyang2022training}. Recent studies \citep{qu2025tabicl,hegselmann2023tabllm} have shown that LLMs can attain competitive performance on tabular classification tasks in few-shot and even zero-shot settings, highlighting their promise for structured climate-disclosure analytics. 

The mainstream LLMs ~\citep{xai_grok3_beta_2025,achiam2023gpt,guo2025deepseek} , built on a decoder-only autoregressive backbone, optimize a straightforward objective (next-word prediction), which is comparatively easy to train and well-suited for generating new tokens~\citep{radford2018improving}.  
Furthermore, the flexibility of the decoder-only model gives the opportunity to perform a branch of different tasks, including text generation (generating rubrics and aggregating rubrics), and quantitative analytics (scoring and evaluation). In our proposed pipeline (Fig.~\ref{fig:our-workflow}), this is realized through the same GPT model. 


\subsection{LLM for scoring/ LLM as a judge}  \label{sec:llm_score}

\begin{figure*}[h]
\centering
\includegraphics[scale=1]{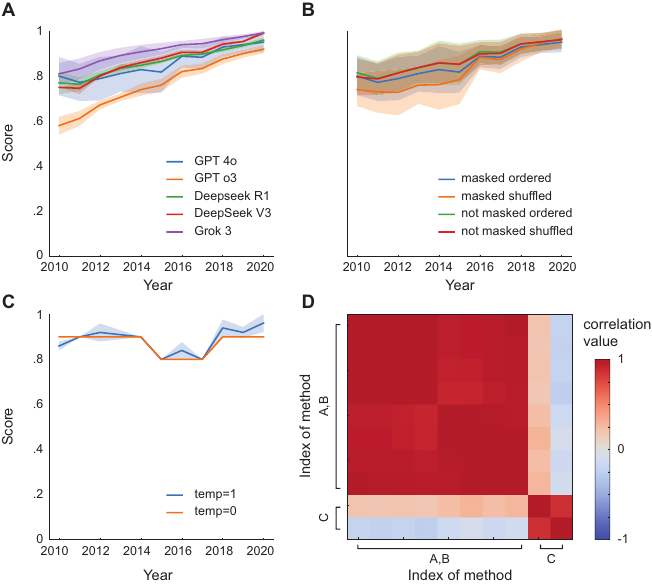}
\caption[n]{\footnotesize 
\textbf{Comparison of different scoring methods.} Bloomberg \textbf{A.} Different large language models (LLM) produce different scoring for the same prompt, but the trend is similar. \textbf{B.} Prompt engineering: hiding explicit years in the questionnaire and/or shuffling the data also gives the same trend. \textbf{C.} When inputting the questionnaire of each year separately into the model, it gives completely different results, and does not show a temporal trend. \textbf{D.} Quantification of the correlation between different methods. 
}
\label{fig:scoring_naive}
\end{figure*}

To test the performance of LLM as a judge, we first tried to na\"ively apply LLMs to grade the question-answer pairs for a representative company (Bloomberg) from 2010 to 2020 using different LLMs (Fig.~\ref{fig:scoring_naive}A) and different prompts (Fig.~\ref{fig:scoring_naive}B). In  Fig.~\ref{fig:scoring_naive}B, we designed the prompt to remove explicit knowledge of the year to ensure that the LLMs give a fair grading. To compare different metrics, we defined the correlation between two time series of scores $s_i(t)$ and $s_j(t)$ as 
\begin{equation}
C_{ij}= \frac{\sum_t\left[ s_i(t) -\langle s_i \rangle\right] \left[ s_j(t) -\langle s_j \rangle\right]}{ \sqrt{\sum_t \left[ s_i(t) -\langle s_i \rangle\right]^2\sum_t \left[ s_j(t) -\langle s_j \rangle\right]^2}},
\label{equ:correlation}
\end{equation}
where $\langle  \cdot \rangle$ represents averaging over time $t$.  
The correlation plot in Fig.~\ref{fig:scoring_naive}D shows all the applied LLMs, and the prompt gives consistent results. It also indicates that Bloomberg's environmental score/awareness increases consistently over time. However, this grading is actually based on cross-year comparison. 
As shown by Fig.~\ref{fig:scoring_naive}C, when using LLM as a judge to grade each question-answer pair individually, the scores are not comparable since they do not have a baseline for comparison~\citep{shen2023large}. This also suggests that without explicit pairwise or groupwise comparison or a set of rules, it is hard to generate reliable scoring aligned to human preference~\citep{ouyang2022training}.
The same phenomenon is observed in~\citep{wei2022chain,wang2022self, wang2022self2}.
To ensure fairness across all companies and years, one would ideally need to present the entire dataset simultaneously and then ask the model to assign scores to each response. However, this approach is infeasible in practice: the CDP dataset is far larger than the limited context window of current LLMs~\citep{touvron2023llama, brown2020language}. Moreover, even if this were technically possible, raw scores alone would yield limited business insight.

\subsection{From individual yearly rubrics to master rubrics}\label{sec:rubric}
\begin{figure}
    \centering
    \includegraphics[width=\linewidth]{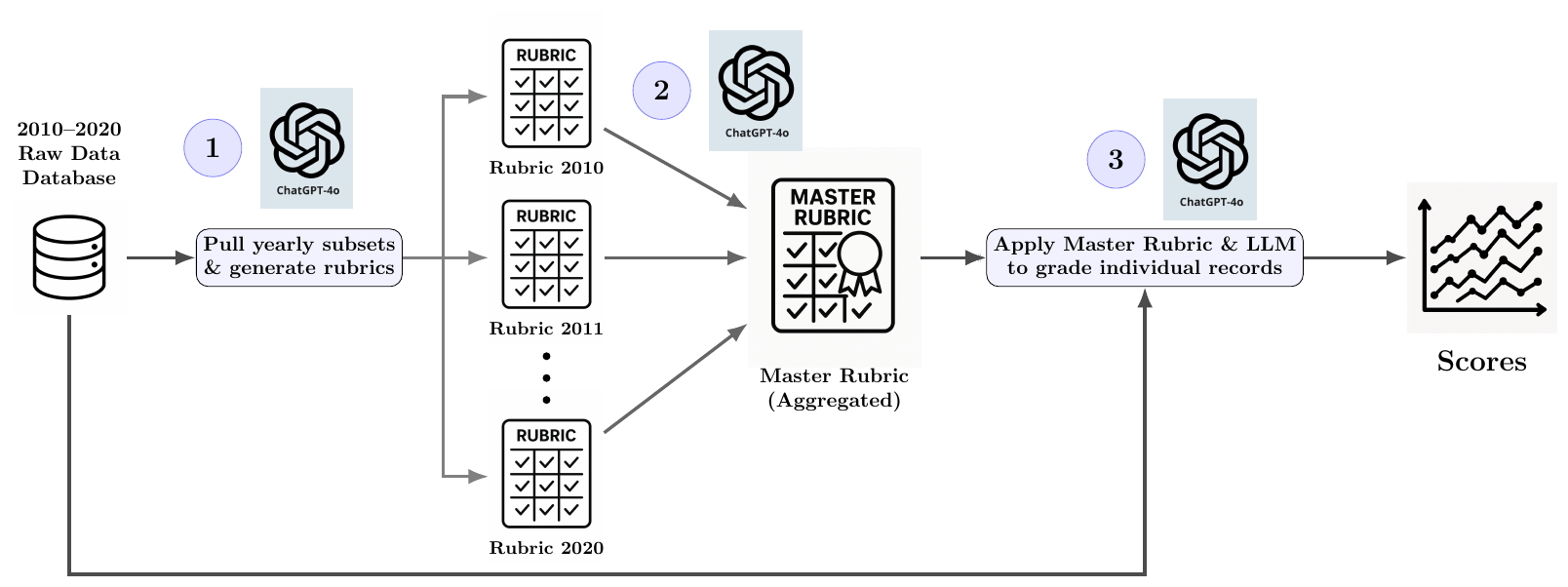}
    \caption{Our main proposed workflow}
    \label{fig:our-workflow}
\end{figure}

A practical alternative is to provide the model with a grading baseline in the form of a rubric to decompose the evaluation criteria~\citep{gu2024survey, li2024llmasajudge}. Early approaches relied on expert-designed rubrics~\citep{liu2023g,kim2023prometheus,kim2024prometheus,zheng2023judging}, while recent advances leverage sample data to automatically induce rubrics, achieving more consistent and interpretable evaluations~\citep{wang2024large, pathak2025rubric}. Although rubric generation still requires exposure to the dataset, this can be addressed through a divide-and-conquer strategy: subsets of the data are input to the model to generate “sub-rubrics”, which are then aggregated. For CDP data, several partitioning strategies are possible. The most straightforward is random sampling, which applies universally but complicates the rubric aggregation step. Alternatively, one can exploit the dataset’s inherent structure. In our case, splitting by company is less effective, since most firms participate only intermittently. Indeed, within the 2010–2020 window we analyze, only $7$ companies participated in every year of CDP reporting. Moreover, splitting by year is most natural: the CDP questionnaire evolves annually due to policy changes, and thus rubrics are not expected to remain stable across years. 

\begin{figure*}[h]
\centering
\includegraphics[scale=1]{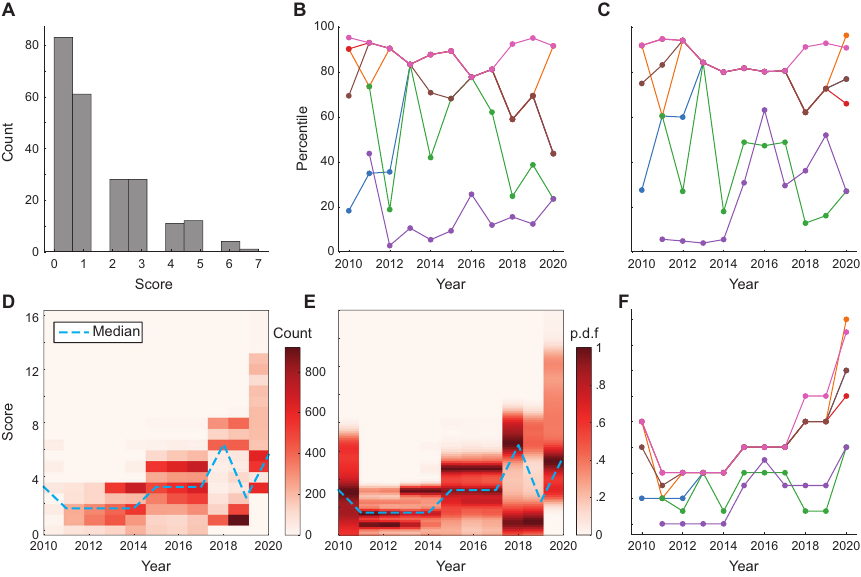}
\caption[n]{\footnotesize 
\textbf{Grading using yearly rubrics and master rubric of CDP data.}  \textbf{A.} Distribution of scores for different companies in the year 2010 using the corresponding yearly rubric. \textbf{B.}, \textbf{C.} Score percentile of example companies over time (\textbf{B.}) using yearly rubrics and (\textbf{C.}) using the master rubric. \textbf{D., E.} Master rubric score of all the companies during 2010-2020 summarized by (\textbf{D.}) count and (\textbf{E.})  probability density function. \textbf{F.} Master rubric score of example companies over time. Example companies in (\textbf{B.}, \textbf{C.}, \textbf{F.}): blue: Tessy plastic, orange: Aptargroup, green: Porton, red: DANFOSS, purple: DOMINGUES PAES EMPRESA DE Segurança, brown: ABM INDUSTRIES Inc., pink: BLOOMBERG LP. 
}
\label{fig:scoring_rubric}
\end{figure*}

\subsubsection{Generation of yearly rubrics}
To generate a yearly rubric, we first input the dataset containing questions and answers on business strategies for all companies within each year separately to LLM (Fig.~\ref{fig:our-workflow}), and let the LLM conclude a rubric that can evaluate the difference among them. We also inquire the LLM to output the scores for each company this year separately. The exact prompt is: ``\textit{First, generate a rubric to evaluate climate responses. Then, based on the rubric to evaluate the answers and output a csv for the companies.}"  At each year, the rubric not only summarizes the questionnaire, but also includes necessary details concluded from the answers of the corresponding year. They contain the rubric for each item (such as ``Strategic Integration" in Table~\ref{tab:2010_rubric}), and corresponding maximum points for each item. 
Moreover, partial credits are given based on the answers. 
For example, the rubric automatically generated by LLM for the year 2010 is in Table~\ref{tab:2010_rubric}. By rubric, the highest score is $10$, but the highest score we observed within all companies is $7$.

\begin{table}[h]
\centering
\scriptsize 
\setlength{\tabcolsep}{4pt} 
\renewcommand{\arraystretch}{1.05} 

\begin{tabularx}{\textwidth}{>{\centering\arraybackslash}m{0.4cm} 
                                  >{\raggedright\arraybackslash}p{3.2cm} 
                                  >{\centering\arraybackslash}m{0.8cm} 
                                  >{\raggedright\arraybackslash}X}
\toprule
\textbf{\#} & \textbf{Item} & \textbf{Max point} & \textbf{Scoring Guidelines} \\
\midrule
\textbf{1} & \textbf{Strategic Integration} & \textbf{2} & 
1 = Mentions climate/sustainability in overall strategy/mission/planning \quad 
2 = Evidence of integration across core functions or competitive advantage \\
\midrule
\textbf{2} & \textbf{Targets \& Metrics} & \textbf{2} & 
1 = Any climate-related target/goal/KPI \quad 
2 = Quantified targets (e.g., \% emissions cut by year) or progress metrics \\
\midrule
\textbf{3} & \textbf{Scenario Analysis} & \textbf{2} & 
1 = References scenario or stress analysis \quad 
2 = Links results explicitly to strategy or financial planning \\
\midrule
\textbf{4} & \textbf{Governance \& Oversight} & \textbf{2} & 
1 = Mentions board/senior management oversight or committees \quad 
2 = Clear roles, reporting lines, or board-level review \\
\midrule
\textbf{5} & \textbf{Stakeholder Engagement \& Disclosure} & \textbf{2} & 
1 = Engages investors/customers/suppliers/policymakers or public communication \quad 
2 = Systematic programs/metrics for engagement or supply-chain collaboration \\
\bottomrule
\end{tabularx}
\caption{Rubric for year 2010. Each item in the rubric contains the maximum points and the meaning of partial and full credits. }
\label{tab:2010_rubric}
\end{table}
The scores based on yearly rubrics are reported in Fig.~\ref{fig:scoring_rubric}A,B. 



\subsubsection{Generation of master rubric} 
To get a fair grading system of question-answer pairs for each year, we further generated a master rubric using the rubric of all years. It is initiated by the prompt: ``\textit{Based on the 11 rubrics generated, generate one comprehensive but concise rubric that works for all rubrics across these years:}" (Fig.~\ref{fig:our-workflow}, step 2). The master rubric we get is presented in Table~\ref{tab:master_rubric}. 

\begin{table}[h]
\centering
\scriptsize 
\setlength{\tabcolsep}{4pt} 
\renewcommand{\arraystretch}{1.05} 
\begin{tabularx}{\textwidth}{>{\centering\arraybackslash}m{0.4cm} 
                                  >{\raggedright\arraybackslash}p{3.6cm} 
                                  >{\centering\arraybackslash}m{0.7cm} 
                                  >{\raggedright\arraybackslash}X}
\toprule
\textbf{\#} & \textbf{Item} & \textbf{Max point} & \textbf{Scoring Guidelines}  \\
\midrule
\textbf{1} & \textbf{Strategic Integration \& Influence} \newline 
\emph{Is climate embedded in core business strategy/financial planning?} 
& \textbf{4} & 
0 = None \quad 
1 = Early-stage/unit mention \quad 
2 = Qualitative/partial integration \quad 
3 = Enterprise-wide with examples \quad 
4 = Shapes enterprise strategy \& long-term financial planning \\
\midrule
\textbf{2} & \textbf{Scenario Analysis} \newline 
\emph{Quality of climate-scenario work and use} 
& \textbf{4} & 
0 = None \quad 
1 = Plan/timeline only \quad 
2 = Qual or quant analysis \quad 
3 = Both, $\geq$1 pathway, weak linkage \quad 
4 = Robust multi-pathway ($\leq$1.5°C, $\geq$2°C) + integrated into strategy/risk \\
\midrule
\textbf{3} & \textbf{Governance \& Oversight} \newline 
\emph{Structures for accountability} 
& \textbf{4} & 
0 = None \quad 
1 = Named body, vague roles \quad 
2 = Defined oversight or policy \quad 
3 = Systematic review + policy/metrics, carbon price planned \quad 
4 = Comprehensive (board mandate, exec incentives, core policies, active carbon price) \\
\midrule
\textbf{4} & \textbf{Targets, Metrics \& Pricing} \newline 
\emph{How performance is measured} 
& \textbf{4} & 
0 = None \quad 
1 = Unquantified goals \quad 
2 = Quantified targets or planned carbon price \quad 
3 = Science-based/time-bound targets or adopted price \quad 
4 = SBTi-aligned targets + KPI disclosure + price shapes investment \\
\midrule
\textbf{5} & \textbf{Disclosure \& Transparency} \newline 
\emph{Depth, specificity \& gaps} 
& \textbf{4} & (Fill in scale as above) \\
\bottomrule
\end{tabularx}
\caption{Master rubric for scoring climate strategy integration and disclosure. Each item in the rubric contains the maximum points and the meaning of partial and full credits. }\label{tab:master_rubric}
\end{table}

After this, we employed the master rubric to grade the question-answer pairs of all companies and years (Fig.~\ref{fig:our-workflow}, step 3). The first thing we need to do is to validate our grading process, and the master rubric captures the essence of the data. We need to confirm that within each year, the ranking among companies is the same using the corresponding individual yearly rubric and master rubric. 
We first chose several sample companies, which appear in most years' datasets. Their scores using yearly rubrics and master rubric are plotted in Fig.~\ref{fig:scoring_rubric}B,C. Most rankings within each year are kept. To quantify this, we calculated Kendall's $\tau$ value, which measures the ordinal association between two measured quantities~\citep{kendall1938new}. It assesses the similarity in the ranking of data when comparing two variables. The coefficient ranges from $-1$ to $+1$, with $+1$ indicating perfect agreement, $-1$ indicating perfect disagreement, and 0 indicating no association. 
\begin{table}[h]
\centering
\scriptsize
\begin{tabular}{cccccccccccc}
\toprule
\textbf{Year} & 2010 & 2011 & 2012 & 2013 & 2014 & 2015 & 2016 & 2017 & 2018 & 2019 & 2020 \\
\textbf{Kendall's $\tau$} & 0.6604 & 0.6814 & 0.5424 & 0.7712 & 0.6125 & 0.5326 & 0.5239 & 0.6013 & 0.8498 & 0.7003 & 0.8327 \\
\bottomrule
\end{tabular}
\caption{Kendall's $\tau$ value between yearly rubric and master rubric per year}
\label{tab:kendaltauu}
\end{table}

For all years, the values are above 0.5 as reported in Table~\ref{tab:kendaltauu}, which indicates that ranks among all companies in each year are at least 75\% preserved. This validates that the master rubric is reproducing the results of the yearly rubric while making the whole dataset comparable. 

In our analysis, besides reporting raw scores for each company by year, we also computed the percentile rank of each score within the given year (see Fig.~\ref{fig:scoring_rubric}B, C). This relative metric enables us to assess how a company’s environmental awareness evolves over time, when compared against its peers. Using percentile-based measures helps normalize across changing absolute levels and mitigates issues tied to skewed score distributions, making it easier to detect relative improvements or declines, even if the global norms shift. Percentile ranking has proven useful in diverse domains—such as environmental indicators and bibliometrics—for enabling fair, relative comparisons across time and entities~\citep{bornmann2013use,kyaw2022effect,boffo2020esg}. In so doing, this method underscores not just absolute progress but the trajectory of environmental awareness relative to others, providing sharper insight into leaders and laggards.

\section{Business insight and policy interpretation of CDP trends}
\label{sec:policyinterpretation}

Understanding the evolution of corporate climate strategies requires not only robust analytical tools but also contextual interpretation of disclosure dynamics across sectors, geographies, and time. In this section, we apply the LLM-based scoring framework developed earlier to extract actionable insights from the CDP dataset, spanning over a decade of environmental disclosures. By aggregating firm-level scores and percentile ranks across sectors and countries, we aim to surface trends that reveal how businesses respond to climate-related policies, investor pressures, and external shocks.

Our analysis focuses on identifying both structural patterns and event-driven shifts in disclosure behavior. We explore which industries lead or lag in strategic climate alignment, how national regulatory frameworks shape disclosure quality, and what temporal trends coincide with major climate policy events such as the Paris Agreement or the COVID-19 pandemic. We further examine inter-sectoral and inter-country correlations to uncover shared trajectories and divergence points in sustainability communication.

Ultimately, this section links the quantitative outputs of our scoring system with qualitative business and policy contexts. The goal is to demonstrate how CDP data, when evaluated through LLM-guided methods, can support strategic decision-making for investors, corporations, and regulators alike.

\subsection{Overview of sectoral and national patterns}

\begin{figure*}[htbp]
\centering
\includegraphics[scale=1]{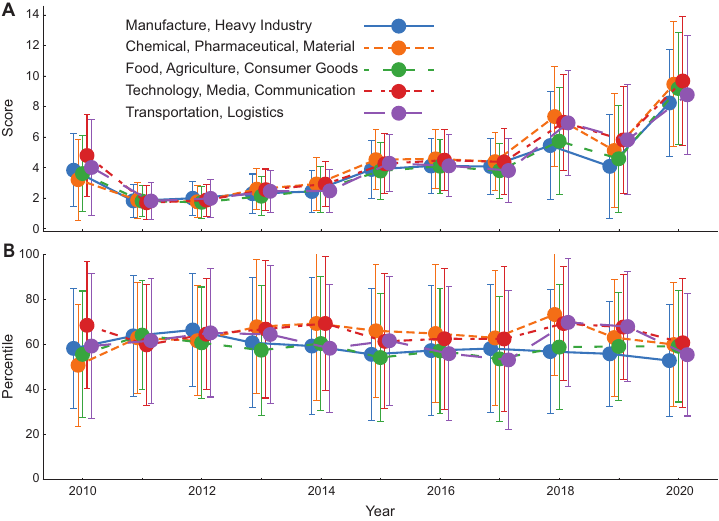}
\caption[n]{\footnotesize 
\textbf{Scores and percentile of companies within each sector based on master rubric. }  \textbf{A.} Average score and \textbf{B.} average percentile for all companies within each sector. The full sector names are: Manufacturing \& Heavy Industry; Chemicals, Pharmaceuticals \& Materials; Food, Agriculture \& Consumer Goods; Technology, Media \& Communications; Transportation \& Logistics. 
}
\label{fig:sector_scores}
\end{figure*}

\begin{figure*}[h]
\centering
\includegraphics[scale=1]{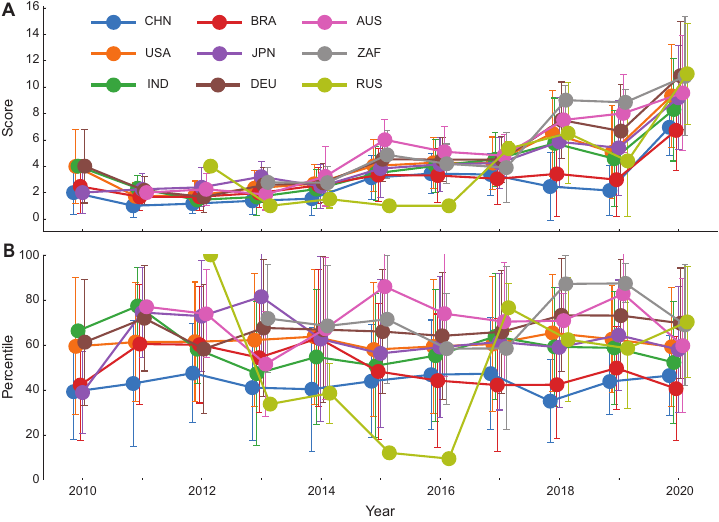}
\caption[n]{\footnotesize 
\textbf{Scores and percentiles of companies within each country based on the master rubric. }  \textbf{A.} Average score and \textbf{B.} average percentile for all companies within each country. 
}
\label{fig:country_comparison}
\end{figure*}

Using the LLM-based scoring framework introduced in Section~\ref{sec:methodology}, we evaluate CDP disclosure quality across sectors and countries from 2010 to 2020. 
The raw scores of most companies show a consistent increasing trend (Fig.~\ref{fig:scoring_rubric}F), which indicates that environmental awareness has an increasing trend across sectors and countries. However, scores do not provide information on how different sectors or countries compare with each other, e.g. Figs.~\ref{fig:sector_scores}A,~\ref{fig:country_comparison}A,~\ref{fig:sector_correlaiton}A,~\ref{fig:country_correlation}A show that all country- or sector-averaged scores increase consistently. To decipher the changes among them, we compared the temporal trend of their percentiles. 
Fig.~\ref{fig:sector_scores} shows that the Technology, Media, and Communication (TMC) sector consistently achieves higher percentile scores, suggesting early adoption of strategic climate integration and clearer reporting standards. In contrast, sectors such as Manufacturing, Transportation, and Heavy Industry lag behind, likely due to the complexity of Scope 3 emissions, cost barriers, and fragmented reporting systems.

At the national level, as shown in Fig.~\ref{fig:country_comparison}, countries like Germany, the United States, and Japan maintain high average disclosure scores across multiple years. This trend likely reflects mature regulatory frameworks, active investor engagement, and industry participation in ESG standard-setting initiatives. Conversely, disclosure quality in China, Brazil, and Russia remains lower, with score volatility indicating inconsistent alignment with international reporting norms.


\subsection{Policy events and disclosure disruptions}

To contextualize the observed score dynamics, we align them with major international climate policy milestones. Table~\ref{tab:policy_events} outlines key global events alongside score trends. Notably, the Paris Agreement (2015) and the IPCC 1.5$^\circ$C report (2018) coincide with increased sectoral engagement and disclosure quality in 2016 and 2019, respectively (Figure \ref{fig:sector_scores}).

However, we also observe sharp declines in disclosure participation and score dispersion around 2017–2018 (Fig.~\ref{fig:country_comparison}), particularly among companies in the United States, China, and Brazil. This may reflect shifting regulatory attitudes, political transitions, or weakening investor mandates. This sharp decline could also possibly be attributed to the nuances and extra workload required in reporting since CDP adopted a new reporting framework in 2018. The COVID-19 pandemic appears to have introduced stagnation in both participation and score movement in 2020, possibly due to operational disruptions and resource reallocation.

\begin{table}[h]
\centering
\small
\begin{tabular}{llp{7.5cm}}
\hline
\textbf{Year} & \textbf{Event} & \textbf{Observed Disclosure Pattern} \\
\hline
2015 & Paris Agreement & Uptick in disclosure scores and participation in 2015, especially in Europe and TMC sectors  \\
2018 & IPCC 1.5°C Report & Renewed strategic language and increased depth in risk-related disclosures \\
2020 & COVID-19 Pandemic & Flat trend in scores and reduced disclosure participation globally \\
2023 & ISSB/IFRS S2 (anticipated) & May drive future harmonization and regulatory benchmarking \\
\hline
\end{tabular}
\caption{Climate Policy Events and Observed CDP Score Trends}
\label{tab:policy_events}
\end{table}

\subsection{Inter-sector and inter-country correlation}

\begin{figure*}[h]
\centering
\includegraphics[scale=1]{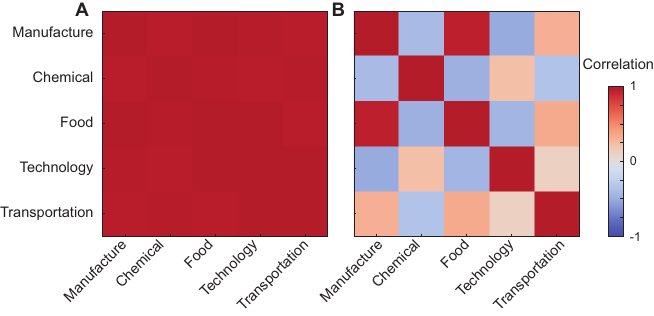}
\caption[n]{\footnotesize 
\textbf{Correlation between scores and percentile of companies between different sectors based on master rubric. }  
Correlation between \textbf{A.} average score and \textbf{B.}  average percentile across different sectors. 
}
\label{fig:sector_correlaiton}
\end{figure*}

To explore cross-domain disclosure behaviors, we compute pairwise correlation matrices between sectoral and country scores over time. 
We calculated the correlation between scores of different sectors and countries using \eqref{equ:correlation}.
Fig.~\ref{fig:sector_correlaiton} reveals high positive correlations between sectors with overlapping regulatory exposure—such as Manufacturing and Food—suggesting parallel adaptation to carbon pricing schemes, supply chain mandates, and environmental taxation. Other pairs of sectors show weak correlation of score over time.

\begin{figure*}[h]
\centering
\includegraphics[scale=1]{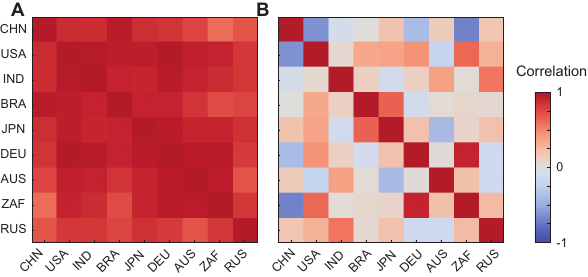}
\caption[n]{\footnotesize 
\textbf{Correlation between scores and percentile of companies between different countries based on master rubric. }  
Correlation between \textbf{A.} average score and \textbf{B.}  average percentile across different countries. 
}
\label{fig:country_correlation}
\end{figure*}

At the national level (Fig.~\ref{fig:country_correlation}), disclosure patterns of Germany \footnote{ We use Germany as a proxy for Europe because it has the largest number of companies in Europe.} shows strong temporal alignment, likely due to synchronized climate policy under the EU ETS and increasing convergence around the Corporate Sustainability Reporting Directive (CSRD). 
In contrast, countries like Brazil and China exhibit more volatile score trajectories, influenced by shifting domestic policy and enforcement capacity.

\subsection{Temporal dynamics in sectoral and national climate disclosures}

To enhance temporal granularity in evaluating disclosure trends, we conduct a year-over-year statistical comparison of both \textbf{average scores} and \textbf{percentile rankings} derived using the master rubric. Figs.~\ref{fig:sector_comparison_pvalue} and~\ref{fig:country_comparison_pvalue} visualize the \textbf{log-transformed inverse p-values} from annual pairwise t-tests, highlighting periods of statistically significant change across \textbf{sectors} and \textbf{countries}, respectively.

\subsubsection{Sectoral dynamics}
Fig.~\ref{fig:sector_comparison_pvalue}A illustrates fluctuations in average climate disclosure scores across industry sectors from 2010 to 2020. The \textit{p-values} reflect statistical confidence in detecting year-to-year changes. Notably:

\begin{figure*}[h]
\centering
\includegraphics[scale=1]{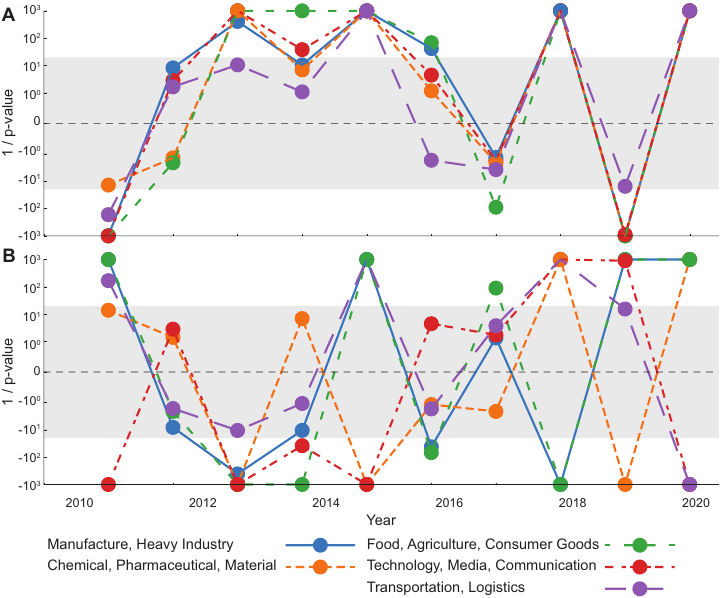}
\caption[n]{\footnotesize 
\textbf{Time evolution of Scores and percentile of companies within each sector based on master rubric.
}  
P-values between each year and the previous year for \textbf{A.} average score and \textbf{B.} average percentile over time. A positive p-value indicates that the score or percentile increases compared to the previous year, vice versa. The grey regime indicates p-value $>0.05$, which indicates that the change is not significant. When p-value $<10^{-3}$, it is cropped to $10^{-3}$ in this plot. 
}
\label{fig:sector_comparison_pvalue}
\end{figure*}

\begin{itemize}
    \item Significant positive shifts occurred following key policy events like the \textit{Paris Agreement (2015)} and \textit{IPCC’s 1.5°C report (2018)}, particularly among \textit{Technology, Media \& Communication (TMC)} and \textit{Chemical sectors}, suggesting heightened responsiveness to international policy signals.
    \item Conversely, \textit{Manufacturing and Transportation sectors} show lower volatility and fewer periods of significant change, reflecting institutional inertia or operational rigidity in sustainability strategy adaptation.
\end{itemize}

Fig.~\ref{fig:sector_comparison_pvalue}B, tracking average \textit{percentile changes}, reveals similar trends with less variance, supporting the hypothesis that disclosure quality growth is not solely due to absolute score improvement but relative benchmarking progress as well. Grey regions denote \textbf{non-significant changes} ($p > 0.05$), while abrupt spikes suggest \textbf{event-driven disclosure shocks}.

\subsubsection{Country-level temporal patterns}

\begin{figure*}[h]
\centering
\includegraphics[scale=1]{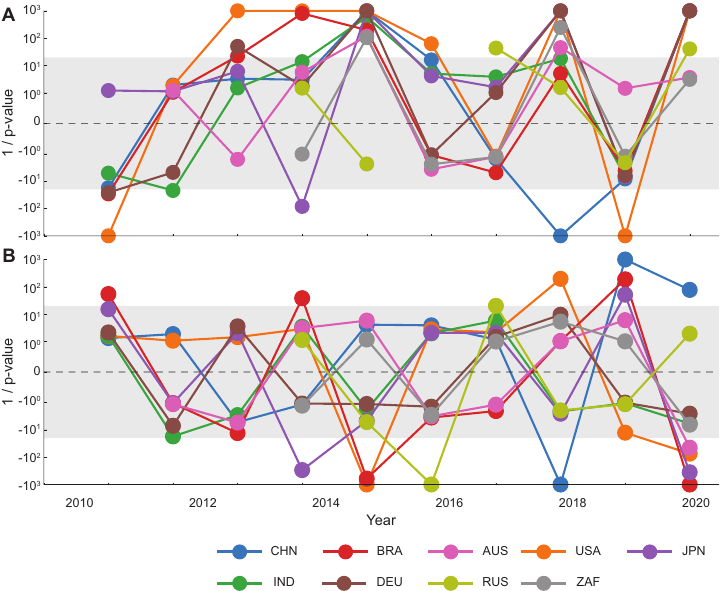}
\caption[n]{\footnotesize 
\textbf{Time evolution of Scores and percentile of companies within each country based on master rubric.
}  
P values between each year and the previous year for \textbf{A.} average score and \textbf{B.} average percentile over time. A positive p-value indicates that the score or percentile increases compared to the previous year, vice versa. The grey regime indicates p value $>0.05$, which indicates that the change is not significant. When p-value $<10^{-3}$, it is cropped to $10^{-3}$ in this plot. 
}
\label{fig:country_comparison_pvalue}
\end{figure*}

Fig.~\ref{fig:country_comparison_pvalue}A presents annual shifts in average scores by country. Western nations such as \textit{Germany, the United States, and Japan} display frequent and significant disclosure adjustments, often aligning with regulatory advancements like \textit{TCFD}, \textit{EU CSRD}, and domestic ESG mandates.

In contrast, developing economies like \textit{Brazil, Russia, and China} exhibit high score volatility but low consistency, reinforcing earlier findings on fragmented regulatory enforcement and potentially strategic or reactionary disclosure patterns.

Fig.~\ref{fig:country_comparison_pvalue}B echoes these trends in percentile-based metrics, reinforcing that changes are not uniformly distributed across countries, even when controlling for baseline disclosure quality. The visualization confirms that LLM-based scoring exposes latent divergence in ESG maturity trajectories and enables \textbf{longitudinal benchmarking}, a key DSS function.

\subsection{Strategic and decision support implications}

The results of our LLM-based scoring and temporal analysis reveal actionable implications for a range of strategic actors within the ESG ecosystem. By systematically quantifying climate strategy integration, governance structures, and disclosure robustness, our master rubric framework offers several key insights that directly inform DSSs:

\textbf{For Investors:} The consistent outperformance of firms in sectors like Technology and regions such as Germany and Japan indicates mature ESG strategies. Investors can use percentile-ranked scores as quantitative signals to screen portfolios for genuine climate commitment. Moreover, the correlation analysis (Figs.~\ref{fig:sector_correlaiton}–\ref{fig:country_correlation}) helps detect sector- and country-level convergence or divergence, offering insights for sector rotation strategies or regional ESG index construction.

\textbf{For Policymakers:} Our results reveal policy transmission effects across borders. For example, post-Paris Agreement (2015) trends show score uplift in EU-aligned countries, indicating the spillover of disclosure standards via regional policy coordination. The temporal p-value analysis (Figs.~\ref{fig:sector_comparison_pvalue}–\ref{fig:country_comparison_pvalue}) can help regulators pinpoint periods of weak alignment or strategic regression—enabling targeted interventions and enforcement.

\textbf{For Corporate Managers:} High-scoring firms can leverage their ESG leadership as a competitive differentiator. Percentile trends (Fig.~\ref{fig:country_comparison_pvalue}) offer a relative benchmark to track progress against industry peers over time. Additionally, companies in lagging sectors (e.g., Manufacturing, Transportation) can use master rubric dimensions to prioritize governance and strategy integration improvements.

\textbf{Cross-Cutting DSS Benefits:} Embedding LLM-based disclosure evaluation into DSS tools enables:
\begin{itemize}
    \item Early warning detection for greenwashing via low rubric alignment (cf.~\citet{chuang2025judging})
    \item Forward planning based on scenario analysis metrics
    \item ESG roadmap benchmarking for internal governance
\end{itemize}

In essence, our framework shifts climate disclosure evaluation from post-hoc narrative analysis to \textit{proactive, decision-aligned strategy assessment}. This reinforces the value of CDP data not only as a reporting mechanism but as an instrument for corporate foresight, regulatory planning, and sustainable investment.

\subsection{Summary of key trends and decision implications}

This section synthesizes the quantitative dynamics of scoring with the qualitative nuances of business strategy and policy developments to extract comprehensive insights into global climate disclosure patterns. By applying our LLM-driven master rubric in conjunction with a percentile-based normalization framework, we identified several key trends. First, there is clear \textbf{sectoral divergence} in disclosure quality—sectors such as TMC and Chemicals show greater integration of climate strategy, while traditional heavy industries exhibit slower adaptation. Second, \textbf{geographic stratification} is evident, with countries like Germany, Japan, and the United States consistently demonstrating higher alignment with international ESG frameworks, in contrast to more variable and less consistent patterns observed in Brazil, China, and Russia. Third, we observe \textbf{event-correlated shifts} in disclosure depth that coincide with major climate milestones such as the Paris Agreement and the IPCC reports, indicating the responsiveness of corporate communication to global regulatory cues. Finally, \textbf{temporal benchmarking} using p-value analysis highlights statistically significant shifts in disclosure quality over time, enabling rigorous year-over-year tracking of ESG narrative evolution. 

Collectively, these findings demonstrate how our framework converts the CDP’s complex and heterogeneous textual disclosures into structured, comparable, and actionable intelligence. This reinforces the broader utility of rubric-guided LLM analytics in supporting strategic ESG planning, regulatory evaluation, and investment decision-making—delivering on the promise of AI-powered sustainability intelligence.







\section{Conclusion}\label{sec:conclusion}
In this paper, we introduced an LLM-based framework for evaluating the quality of corporate climate disclosures submitted to the CDP over the 2010--2020 period. Our goal was to demonstrate how LLMs---beyond their conversational applications---can support structured decision-making in climate governance by processing large volumes of unstructured textual data. To this end, we designed a rubric-guided scoring pipeline that leverages the reasoning capabilities of instruction-tuned LLMs. Our approach incorporates both year-specific and harmonized evaluation criteria, enabling comparability over time despite shifting disclosure formats and policy contexts. The resulting dataset, comprising nearly 20{,}000 company-year-level scores across five interpretive dimensions, provides a consistent lens through which to assess disclosure quality, strategy integration, and ESG maturity across sectors and geographies.

Our findings reveal several meaningful trends. While early-year responses often emphasized vague commitments or general awareness, later disclosures increasingly incorporated specific metrics, climate governance structures, and references to scenario planning. We also observed heterogeneity in disclosure quality both across sectors and within the same year, suggesting that regulatory alignment, peer benchmarking, and external pressure continue to shape how companies communicate their climate strategy. Importantly, the scores generated by our method correlate with established policy milestones, underscoring the utility of LLMs in interpreting time-sensitive ESG signals.

This study contributes to the Decision Support literature in three important ways. First, it expands the toolkit for decision-makers seeking scalable methods to extract structured insights from open-ended disclosures. Second, it demonstrates how LLMs can function as evaluative agents capable of applying contextual rubrics and tracking semantic consistency over time. Third, it introduces a publicly usable dataset and scoring template that can be extended, replicated, or integrated into ESG dashboards and supply chain assessment tools.

The work also opens several directions for future research. One avenue is integrating LLM-generated disclosure scores with structured Compustat financial data to investigate how narrative climate quality correlates with firm-level financial outcomes such as revenue growth, cost of capital, or volatility. This merger could reveal whether stronger climate communication translates into increased investor confidence, lower ESG risk premiums, or greater resilience during crises such as COVID-19 or energy shocks. Another avenue is enriching the analysis with Scope~1, 2, and 3 greenhouse gas emissions data from structured CDP fields, which would allow for triangulation between stated intent and actual performance. This approach could distinguish companies with aligned narratives and genuine decarbonization trajectories from those with superficial or lagging commitments, while also enabling comparisons of sector-specific reporting norms.

A third direction is to extend the framework to investor-requested CDP disclosures, which are often more detailed and shareholder-oriented, reflecting different incentives and regulatory pressures. Applying the rubric to these responses could uncover how disclosure tone and quality shift under investor scrutiny. Combining this with data on investor behavior, such as ESG voting records or fund flows, could lead to the development of new climate risk signaling models. 

Finally, applying semantic embeddings derived from LLMs, including sentence- or document-level vector representations, offers deeper analytical capabilities. Embeddings could be used to cluster disclosures by strategic themes or ESG maturity stages, conduct semantic similarity searches to identify peer companies or exemplar disclosures, and support transfer learning where pretrained embeddings on climate texts are adapted to other ESG domains such as biodiversity, diversity--equity--inclusion (DEI), or just transition. They also enable anomaly detection to flag outliers in tone, coverage, or ambition, and support visualization of disclosure evolution over time using methods such as UMAP or t-SNE, thereby allowing intuitive exploration of how companies converge or diverge in climate strategy space.

\bibliographystyle{abbrvnat}
\bibliography{reference}

\begin{thebibliography}{60}
\providecommand{\natexlab}[1]{#1}
\providecommand{\url}[1]{\texttt{#1}}
\expandafter\ifx\csname urlstyle\endcsname\relax
  \providecommand{\doi}[1]{doi: #1}\else
  \providecommand{\doi}{doi: \begingroup \urlstyle{rm}\Url}\fi

\bibitem[Achiam et~al.(2023)Achiam, Adler, Agarwal, Ahmad, Akkaya, Aleman, Almeida, Altenschmidt, Altman, Anadkat, et~al.]{achiam2023gpt}
J.~Achiam, S.~Adler, S.~Agarwal, L.~Ahmad, I.~Akkaya, F.~L. Aleman, D.~Almeida, J.~Altenschmidt, S.~Altman, S.~Anadkat, et~al.
\newblock Gpt-4 technical report.
\newblock \emph{arXiv preprint arXiv:2303.08774}, 2023.

\bibitem[Armbrust(2022)]{armbrust2022deep}
F.~Armbrust.
\newblock \emph{Deep Sustainable Finance: An End-to-End Text Analysis of the Financial and Environmental Narratives in Corporate Disclosures}.
\newblock PhD thesis, University of Stuttgart, 2022.
\newblock URL \url{https://elib.uni-stuttgart.de/bitstream/11682/12287/3/Armbrust_Felix_Dissertation_2022.pdf}.

\bibitem[Bingler et~al.(2024)Bingler, Kraus, Leippold, and Webersinke]{bingler2024cheap}
J.~Bingler, M.~Kraus, M.~Leippold, and N.~Webersinke.
\newblock How cheap talk in climate disclosures relates to climate initiatives, corporate emissions, and reputation risk.
\newblock \emph{SSRN}, 2024.
\newblock URL \url{https://papers.ssrn.com/sol3/papers.cfm?abstract_id=4000708}.

\bibitem[Blanco(2021)]{blanco2021supply}
C.~C. Blanco.
\newblock Supply chain carbon footprinting and climate change disclosures of global firms.
\newblock \emph{Production and Operations Management}, 30\penalty0 (9):\penalty0 3143--3160, 2021.

\bibitem[Blei et~al.(2003)Blei, Ng, and Jordan]{blei2003latent}
D.~M. Blei, A.~Y. Ng, and M.~I. Jordan.
\newblock Latent dirichlet allocation.
\newblock \emph{Journal of machine Learning research}, 3\penalty0 (Jan):\penalty0 993--1022, 2003.

\bibitem[Boffo et~al.(2020)Boffo, Marshall, and Patalano]{boffo2020esg}
R.~Boffo, C.~Marshall, and R.~Patalano.
\newblock Esg investing: Environmental pillar scoring and reporting.
\newblock \emph{Retrived}, 14:\penalty0 2021, 2020.

\bibitem[Bornmann et~al.(2013)Bornmann, Leydesdorff, and Mutz]{bornmann2013use}
L.~Bornmann, L.~Leydesdorff, and R.~Mutz.
\newblock The use of percentiles and percentile rank classes in the analysis of bibliometric data: Opportunities and limits.
\newblock \emph{Journal of informetrics}, 7\penalty0 (1):\penalty0 158--165, 2013.

\bibitem[Brown et~al.(2020)Brown, Mann, Ryder, Subbiah, Kaplan, Dhariwal, Neelakantan, Shyam, Sastry, Askell, et~al.]{brown2020language}
T.~Brown, B.~Mann, N.~Ryder, M.~Subbiah, J.~D. Kaplan, P.~Dhariwal, A.~Neelakantan, P.~Shyam, G.~Sastry, A.~Askell, et~al.
\newblock Language models are few-shot learners.
\newblock \emph{Advances in neural information processing systems}, 33:\penalty0 1877--1901, 2020.

\bibitem[Cao et~al.(2025)Cao, Han, Wang, and Jia]{cao2025carbonchat}
Z.~Cao, M.~Han, J.~Wang, and M.~Jia.
\newblock Carbonchat: Large language model-based corporate carbon emission analysis and climate knowledge q\&a system.
\newblock \emph{arXiv preprint arXiv:2501.02031}, 2025.
\newblock URL \url{https://arxiv.org/abs/2501.02031}.

\bibitem[Chuang et~al.(2025{\natexlab{a}})Chuang, Chuang, and Chuang]{chuang2025judging}
M.~Chuang, G.~Chuang, and J.~Chuang.
\newblock Judging it, washing it: Scoring and greenwashing corporate climate disclosures using large language models.
\newblock \emph{arXiv preprint arXiv:2502.15094}, 2025{\natexlab{a}}.
\newblock URL \url{https://arxiv.org/abs/2502.15094}.

\bibitem[Chuang et~al.(2025{\natexlab{b}})]{Chuang2025Greenwashing}
Y.~Chuang et~al.
\newblock Judging it, washing it: Scoring and greenwashing corporate climate disclosures using large language models.
\newblock \emph{Environmental Science \& Technology}, 2025{\natexlab{b}}.

\bibitem[Climate(2023)]{Matisoff2023}
M.~Climate.
\newblock Cdp reporting annual disclosure updates for 2023., 2023.
\newblock URL \url{https://www.manifestclimate.com/blog/cdp-disclosure-updates/#:~:text=The%20question%20under%20%E2%80%9CBusiness%20Strategy%E2%80%9D,affect%20their%20financial%20planning%20processes}.
\newblock website.

\bibitem[Cohen et~al.(2023)Cohen, Kadach, and Ormazabal]{cohen2023institutional}
S.~Cohen, I.~Kadach, and G.~Ormazabal.
\newblock Institutional investors, climate disclosure, and carbon emissions.
\newblock \emph{SSRN}, 2023.
\newblock URL \url{https://papers.ssrn.com/sol3/papers.cfm?abstract_id=4138869}.

\bibitem[Cortes and Vapnik(1995)]{cortes1995support}
C.~Cortes and V.~Vapnik.
\newblock Support-vector networks.
\newblock \emph{Machine learning}, 20\penalty0 (3):\penalty0 273--297, 1995.

\bibitem[Devlin et~al.(2019)Devlin, Chang, Lee, and Toutanova]{devlin2019bert}
J.~Devlin, M.-W. Chang, K.~Lee, and K.~Toutanova.
\newblock Bert: Pre-training of deep bidirectional transformers for language understanding.
\newblock In \emph{Proceedings of the 2019 conference of the North American chapter of the association for computational linguistics: human language technologies, volume 1 (long and short papers)}, pages 4171--4186, 2019.
\newblock URL \url{https://aclanthology.org/N19-1423/}.

\bibitem[D’Amico et~al.(2016)D’Amico, Coluccia, Fontana, and Solimene]{DAmico2016}
E.~D’Amico, D.~Coluccia, S.~Fontana, and S.~Solimene.
\newblock Factors influencing corporate environmental disclosure.
\newblock \emph{Business Strategy and the Environment}, 25\penalty0 (3):\penalty0 178--192, 2016.

\bibitem[Es et~al.(2024)Es, James, Anke, and Schockaert]{es2024ragas}
S.~Es, J.~James, L.~E. Anke, and S.~Schockaert.
\newblock Ragas: Automated evaluation of retrieval augmented generation.
\newblock In \emph{Proceedings of the 18th Conference of the European Chapter of the Association for Computational Linguistics: System Demonstrations}, pages 150--158, 2024.

\bibitem[Fang et~al.(2024)Fang, Xu, Tan, Zhang, Hu, Qi, Nickleach, Socolinsky, Sengamedu, and Faloutsos]{fang2024large}
X.~Fang, W.~Xu, F.~A. Tan, J.~Zhang, Z.~Hu, Y.~Qi, S.~Nickleach, D.~Socolinsky, S.~Sengamedu, and C.~Faloutsos.
\newblock Large language models (llms) on tabular data: Prediction, generation, and understanding--a survey.
\newblock \emph{arXiv preprint arXiv:2402.17944}, 2024.

\bibitem[Gu et~al.(2024)Gu, Jiang, Shi, Tan, Zhai, Xu, Li, Shen, Ma, Liu, et~al.]{gu2024survey}
J.~Gu, X.~Jiang, Z.~Shi, H.~Tan, X.~Zhai, C.~Xu, W.~Li, Y.~Shen, S.~Ma, H.~Liu, et~al.
\newblock A survey on llm-as-a-judge.
\newblock \emph{arXiv preprint arXiv:2411.15594}, 2024.

\bibitem[Guo et~al.(2025)Guo, Yang, Zhang, Song, Zhang, Xu, Zhu, Ma, Wang, Bi, et~al.]{guo2025deepseek}
D.~Guo, D.~Yang, H.~Zhang, J.~Song, R.~Zhang, R.~Xu, Q.~Zhu, S.~Ma, P.~Wang, X.~Bi, et~al.
\newblock Deepseek-r1: Incentivizing reasoning capability in llms via reinforcement learning.
\newblock \emph{arXiv preprint arXiv:2501.12948}, 2025.

\bibitem[Gweon and Schonlau(2023)]{gweon2023automated}
H.~Gweon and M.~Schonlau.
\newblock Automated classification for open-ended questions with bert.
\newblock \emph{arXiv preprint arXiv:2209.06178}, 2023.

\bibitem[Haghighat et~al.()Haghighat, Dao, Qader, Dubayah, and Kiely]{haghighat2025sota}
A.~Haghighat, T.~Dao, A.~Qader, B.~Dubayah, and P.~Kiely.
\newblock How we run gpt oss 120b at 500+ tokens per second on nvidia gpus.
\newblock Baseten Blog.
\newblock URL \url{https://www.baseten.co/blog/sota-performance-for-gpt-oss-120b-on-nvidia-gpus/}.
\newblock Last updated August 7, 2025.

\bibitem[Hegselmann et~al.(2023)Hegselmann, Buendia, Lang, Agrawal, Jiang, and Sontag]{hegselmann2023tabllm}
S.~Hegselmann, A.~Buendia, H.~Lang, M.~Agrawal, X.~Jiang, and D.~Sontag.
\newblock Tabllm: Few-shot classification of tabular data with large language models.
\newblock In \emph{International conference on artificial intelligence and statistics}, pages 5549--5581. PMLR, 2023.

\bibitem[Kendall(1938)]{kendall1938new}
M.~G. Kendall.
\newblock A new measure of rank correlation.
\newblock \emph{Biometrika}, 30\penalty0 (1-2):\penalty0 81--93, 1938.

\bibitem[Kim et~al.(2023)Kim, Shin, Cho, Jang, Longpre, Lee, Yun, Shin, Kim, Thorne, et~al.]{kim2023prometheus}
S.~Kim, J.~Shin, Y.~Cho, J.~Jang, S.~Longpre, H.~Lee, S.~Yun, S.~Shin, S.~Kim, J.~Thorne, et~al.
\newblock Prometheus: Inducing fine-grained evaluation capability in language models.
\newblock In \emph{The Twelfth International Conference on Learning Representations}, 2023.

\bibitem[Kim et~al.(2024)Kim, Suk, Longpre, Lin, Shin, Welleck, Neubig, Lee, Lee, and Seo]{kim2024prometheus}
S.~Kim, J.~Suk, S.~Longpre, B.~Y. Lin, J.~Shin, S.~Welleck, G.~Neubig, M.~Lee, K.~Lee, and M.~Seo.
\newblock Prometheus 2: An open source language model specialized in evaluating other language models.
\newblock \emph{arXiv preprint arXiv:2405.01535}, 2024.

\bibitem[Kyaw(2022)]{kyaw2022effect}
K.~Kyaw.
\newblock Effect of policy uncertainty on environmental innovation.
\newblock \emph{Journal of Cleaner Production}, 363:\penalty0 132645, 2022.

\bibitem[Lee et~al.(2024)]{Lee2024CoT}
S.~Lee et~al.
\newblock Applying large language models and chain-of-thought for automatic scoring.
\newblock \emph{arXiv preprint arXiv:2312.03748}, 2024.
\newblock URL \url{https://arxiv.org/abs/2312.03748}.

\bibitem[Leng et~al.(2023)Leng, Uhlenhuth, and Polyzotis]{Leng2023}
Q.~Leng, K.~Uhlenhuth, and A.~Polyzotis.
\newblock Best practices for llm evaluation of rag applications, 2023.
\newblock URL \url{https://www.databricks.com/blog/LLM-auto-eval-best-practices-RAG}.
\newblock Retrieved from Databricks Blog.

\bibitem[Leviathan et~al.(2023)Leviathan, Kalman, and Matias]{leviathan2023fast}
Y.~Leviathan, M.~Kalman, and Y.~Matias.
\newblock Fast inference from transformers via speculative decoding.
\newblock In \emph{International Conference on Machine Learning}, pages 19274--19286. PMLR, 2023.

\bibitem[Li et~al.(2024)Li, Jiang, Huang, Beigi, Zhao, Tan, Bhattacharjee, Jiang, Chen, Wu, Shu, Cheng, and Liu]{li2024llmasajudge}
D.~Li, B.~Jiang, L.~Huang, A.~Beigi, C.~Zhao, Z.~Tan, A.~Bhattacharjee, Y.~Jiang, C.~Chen, T.~Wu, K.~Shu, L.~Cheng, and H.~Liu.
\newblock From generation to judgment: Opportunities and challenges of llm-as-a-judge.
\newblock \emph{arXiv preprint arXiv: 2411.16594}, 2024.

\bibitem[Li et~al.(2010)]{li2010textual}
F.~Li et~al.
\newblock Textual analysis of corporate disclosures: A survey of the literature.
\newblock \emph{Journal of accounting literature}, 29\penalty0 (1):\penalty0 143--165, 2010.

\bibitem[Liu et~al.(2024)Liu, Feng, Xue, Wang, Wu, Lu, Zhao, Deng, Zhang, Ruan, et~al.]{liu2024deepseek}
A.~Liu, B.~Feng, B.~Xue, B.~Wang, B.~Wu, C.~Lu, C.~Zhao, C.~Deng, C.~Zhang, C.~Ruan, et~al.
\newblock Deepseek-v3 technical report.
\newblock \emph{arXiv preprint arXiv:2412.19437}, 2024.

\bibitem[Liu et~al.(2019)Liu, Ott, Goyal, Du, Joshi, Chen, Levy, Lewis, Zettlemoyer, and Stoyanov]{liu2019roberta}
Y.~Liu, M.~Ott, N.~Goyal, J.~Du, M.~Joshi, D.~Chen, O.~Levy, M.~Lewis, L.~Zettlemoyer, and V.~Stoyanov.
\newblock Roberta: A robustly optimized bert pretraining approach.
\newblock \emph{arXiv preprint arXiv:1907.11692}, 2019.

\bibitem[Liu et~al.(2023)Liu, Iter, Xu, Wang, Xu, and Zhu]{liu2023g}
Y.~Liu, D.~Iter, Y.~Xu, S.~Wang, R.~Xu, and C.~Zhu.
\newblock G-eval: Nlg evaluation using gpt-4 with better human alignment.
\newblock \emph{arXiv preprint arXiv:2303.16634}, 2023.

\bibitem[Matisoff et~al.(2013)Matisoff, Noonan, and O'Brien]{Matisoff2013}
D.~C. Matisoff, D.~S. Noonan, and J.~J. O'Brien.
\newblock Convergence in environmental reporting: Assessing the carbon disclosure project.
\newblock \emph{Business Strategy and the Environment}, 22\penalty0 (5):\penalty0 285--305, 2013.

\bibitem[Matsumura et~al.(2014)Matsumura, Prakash, and Vera-Mu{\~n}oz]{matsumura2014firm}
E.~M. Matsumura, R.~Prakash, and S.~C. Vera-Mu{\~n}oz.
\newblock Firm-value effects of carbon emissions and carbon disclosures.
\newblock \emph{The accounting review}, 89\penalty0 (2):\penalty0 695--724, 2014.

\bibitem[Mellon et~al.(2024)Mellon, Bailey, Scott, Breckwoldt, Miori, and Schmedeman]{mellon2024important}
J.~Mellon, J.~Bailey, R.~Scott, J.~Breckwoldt, M.~Miori, and P.~Schmedeman.
\newblock Do ais know what the most important issue is? using language models to code open-text social survey responses at scale.
\newblock \emph{SSRN}, 2024.
\newblock URL \url{https://papers.ssrn.com/sol3/papers.cfm?abstract_id=4310154}.

\bibitem[Mizumoto and Eguchi(2023)]{Mizumoto2023}
A.~Mizumoto and M.~Eguchi.
\newblock Applying large language models for automated essay scoring.
\newblock \emph{Language Assessment Quarterly}, 2023.

\bibitem[Ouyang et~al.(2022)Ouyang, Wu, Jiang, Almeida, Wainwright, Mishkin, Zhang, Agarwal, Slama, Ray, et~al.]{ouyang2022training}
L.~Ouyang, J.~Wu, X.~Jiang, D.~Almeida, C.~Wainwright, P.~Mishkin, C.~Zhang, S.~Agarwal, K.~Slama, A.~Ray, et~al.
\newblock Training language models to follow instructions with human feedback.
\newblock \emph{Advances in neural information processing systems}, 35:\penalty0 27730--27744, 2022.

\bibitem[Pathak et~al.(2025)Pathak, Gandhi, Uttam, Ramamoorthy, Ghosh, Jindal, Verma, Mittal, Ased, Khatri, et~al.]{pathak2025rubric}
A.~Pathak, R.~Gandhi, V.~Uttam, A.~Ramamoorthy, P.~Ghosh, A.~R. Jindal, S.~Verma, A.~Mittal, A.~Ased, C.~Khatri, et~al.
\newblock Rubric is all you need: Enhancing llm-based code evaluation with question-specific rubrics.
\newblock \emph{arXiv preprint arXiv:2503.23989}, 2025.

\bibitem[Petukhova et~al.(2024)]{petukhova2024text}
M.~Petukhova et~al.
\newblock Text clustering with large language model embeddings.
\newblock \emph{arXiv preprint arXiv:2403.15112}, 2024.

\bibitem[Qian et~al.(2024)Qian, Gonugondla, Ha, Shang, Gouda, Nallapati, Sengupta, Ma, and Deoras]{qian2024bass}
H.~Qian, S.~K. Gonugondla, S.~Ha, M.~Shang, S.~K. Gouda, R.~Nallapati, S.~Sengupta, X.~Ma, and A.~Deoras.
\newblock Bass: Batched attention-optimized speculative sampling.
\newblock \emph{arXiv preprint arXiv:2404.15778}, 2024.

\bibitem[Qu et~al.(2025)Qu, Holzm{\~A}{\v{z}}ller, Varoquaux, and Morvan]{qu2025tabicl}
J.~Qu, D.~Holzm{\~A}{\v{z}}ller, G.~Varoquaux, and M.~L. Morvan.
\newblock Tabicl: A tabular foundation model for in-context learning on large data.
\newblock \emph{arXiv preprint arXiv:2502.05564}, 2025.

\bibitem[Radford et~al.(2018)Radford, Narasimhan, Salimans, Sutskever, et~al.]{radford2018improving}
A.~Radford, K.~Narasimhan, T.~Salimans, I.~Sutskever, et~al.
\newblock Improving language understanding by generative pre-training.
\newblock 2018.

\bibitem[Shannon(1948)]{shannon1948mathematical}
C.~E. Shannon.
\newblock A mathematical theory of communication.
\newblock \emph{The Bell system technical journal}, 27\penalty0 (3):\penalty0 379--423, 1948.

\bibitem[Shen et~al.(2023)Shen, Cheng, Nguyen, You, and Bing]{shen2023large}
C.~Shen, L.~Cheng, X.-P. Nguyen, Y.~You, and L.~Bing.
\newblock Large language models are not yet human-level evaluators for abstractive summarization.
\newblock \emph{arXiv preprint arXiv:2305.13091}, 2023.

\bibitem[Sojasingarayar(2025)]{Niasojasingarayar_2025}
A.~Sojasingarayar.
\newblock Top: Llm/rag evaluation framework, August 2025.
\newblock URL \url{https://www.linkedin.com/posts/aboniasojasingarayar_top-llmrag-evaluation-framework-each-activity-7173233812141772800-S0KX/}.
\newblock LinkedIn post.

\bibitem[Sparck~Jones(1972)]{sparck1972statistical}
K.~Sparck~Jones.
\newblock A statistical interpretation of term specificity and its application in retrieval.
\newblock \emph{Journal of documentation}, 28\penalty0 (1):\penalty0 11--21, 1972.

\bibitem[Team et~al.(2023)Team, Anil, Borgeaud, Alayrac, Yu, Soricut, Schalkwyk, Dai, Hauth, Millican, et~al.]{team2023gemini}
G.~Team, R.~Anil, S.~Borgeaud, J.-B. Alayrac, J.~Yu, R.~Soricut, J.~Schalkwyk, A.~M. Dai, A.~Hauth, K.~Millican, et~al.
\newblock Gemini: a family of highly capable multimodal models.
\newblock \emph{arXiv preprint arXiv:2312.11805}, 2023.

\bibitem[Tidy et~al.(2016)Tidy, Wang, and Hall]{tidy2016role}
M.~Tidy, X.~Wang, and M.~Hall.
\newblock The role of supplier relationship management in reducing greenhouse gas emissions from food supply chains: supplier engagement in the uk supermarket sector.
\newblock \emph{Journal of Cleaner Production}, 112:\penalty0 3294--3305, 2016.

\bibitem[Touvron et~al.(2023)Touvron, Lavril, Izacard, Martinet, Lachaux, Lacroix, Rozi{\`e}re, Goyal, Hambro, Azhar, et~al.]{touvron2023llama}
H.~Touvron, T.~Lavril, G.~Izacard, X.~Martinet, M.-A. Lachaux, T.~Lacroix, B.~Rozi{\`e}re, N.~Goyal, E.~Hambro, F.~Azhar, et~al.
\newblock Llama: Open and efficient foundation language models.
\newblock \emph{arXiv preprint arXiv:2302.13971}, 2023.

\bibitem[Wang et~al.(2022{\natexlab{a}})Wang, Wei, Schuurmans, Le, Chi, Narang, Chowdhery, and Zhou]{wang2022self}
X.~Wang, J.~Wei, D.~Schuurmans, Q.~Le, E.~Chi, S.~Narang, A.~Chowdhery, and D.~Zhou.
\newblock Self-consistency improves chain of thought reasoning in language models.
\newblock \emph{arXiv preprint arXiv:2203.11171}, 2022{\natexlab{a}}.

\bibitem[Wang et~al.(2022{\natexlab{b}})Wang, Kordi, Mishra, Liu, Smith, Khashabi, and Hajishirzi]{wang2022self2}
Y.~Wang, Y.~Kordi, S.~Mishra, A.~Liu, N.~A. Smith, D.~Khashabi, and H.~Hajishirzi.
\newblock Self-instruct: Aligning language models with self-generated instructions.
\newblock \emph{arXiv preprint arXiv:2212.10560}, 2022{\natexlab{b}}.

\bibitem[Wang et~al.(2024)Wang, Wu, Wu, Tao, and Fang]{wang2024large}
Y.~Wang, X.~Wu, H.-T. Wu, Z.~Tao, and Y.~Fang.
\newblock Do large language models rank fairly? an empirical study on the fairness of llms as rankers.
\newblock \emph{arXiv preprint arXiv:2404.03192}, 2024.

\bibitem[Wei et~al.(2022)Wei, Wang, Schuurmans, Bosma, Xia, Chi, Le, Zhou, et~al.]{wei2022chain}
J.~Wei, X.~Wang, D.~Schuurmans, M.~Bosma, F.~Xia, E.~Chi, Q.~V. Le, D.~Zhou, et~al.
\newblock Chain-of-thought prompting elicits reasoning in large language models.
\newblock \emph{Advances in neural information processing systems}, 35:\penalty0 24824--24837, 2022.

\bibitem[Wen et~al.(2024)Wen, Zhang, Zheng, Xu, and Bian]{wen2024supervised}
X.~Wen, H.~Zhang, S.~Zheng, W.~Xu, and J.~Bian.
\newblock From supervised to generative: A novel paradigm for tabular deep learning with large language models.
\newblock In \emph{Proceedings of the 30th ACM SIGKDD Conference on Knowledge Discovery and Data Mining}, pages 3323--3333, 2024.

\bibitem[{xAI}(2025)]{xai_grok3_beta_2025}
{xAI}.
\newblock {Grok 3 Beta — The Age of Reasoning Agents}.
\newblock News release, xAI website, Feb. 2025.
\newblock Accessed via https://x.ai/news/grok-3 on September 4, 2025.

\bibitem[Zhang(2024)]{Zhang2024AES}
L.~Zhang.
\newblock Scoring corporate climate disclosures via rubric-based prompting with llms.
\newblock \emph{Advances in Environmental Sciences}, 2024.
\newblock Forthcoming.

\bibitem[Zheng et~al.(2023)Zheng, Chiang, Sheng, Zhuang, Wu, Zhuang, Lin, Li, Li, Xing, et~al.]{zheng2023judging}
L.~Zheng, W.-L. Chiang, Y.~Sheng, S.~Zhuang, Z.~Wu, Y.~Zhuang, Z.~Lin, Z.~Li, D.~Li, E.~Xing, et~al.
\newblock Judging llm-as-a-judge with mt-bench and chatbot arena.
\newblock \emph{Advances in neural information processing systems}, 36:\penalty0 46595--46623, 2023.

\end{thebibliography}

\end{document}